\documentclass[aip,pop,preprint]{revtex4-1}%
\usepackage{amssymb,amsfonts,amsmath}
\usepackage{graphicx}
\usepackage{epstopdf}
\usepackage{diagbox}
\usepackage{multirow}
\usepackage{bm}
\usepackage{xcolor}
\usepackage[colorlinks=true,linkcolor=blue]{hyperref}
\usepackage{notes2bib}
\providecommand{\U}[1]{\protect\rule{.1in}{.1in}}
\begin{document}
\title{Time-resolved spectrum output from a grating spectrometer}
\author{Jian Zheng}
\email{jzheng@ustc.edu.cn}
\affiliation{School of Physical Sciences, University of Science and Technology of China,
Hefei, Anhui 230026, P. R. China}
\affiliation{IFSA Collaborative Innovation Center, Shanghai Jiao Tong University, Shanghai 200240, P. R. China}
\affiliation{CAS Center for Excellence in Ultra-intense Laser Science，Shanghai 201800，P. R. China}
\author{Yao-Yuan Liu}
\affiliation{School of Physical Sciences, University of Science and Technology of China,
Hefei, Anhui 230026, P. R. China}

\begin{abstract}
Time-resolved spectra are often recorded in optical Thomson scattering
experiments of laser-produced plasmas. In this essay, the meaning of time-resolved spectra output from a grating spectrometer is examined. Our results show that the recorded signal is indeed the convolution of the response function of dispersion element and the product of instant local dynamic form factor and electron density. 
\end{abstract}

\maketitle

\section{Introduction}

The Thomson scattering is a powerful diagnostics for plasma physics 
because it can provide accurate and reliable information of high 
temperature plasmas \cite{Froula2011book}. In the fields of 
high-energy-density physics in relevance to laser fusion and laboratory 
astrophysics, the Thomson scattering systems have been developed for various 
laser facilities \cite{Fontaine1994, Glenzer1997a, Bai2001, Wang2005, Ross2006, Ross2011, Gong2015, Ross2016, Zhao2018}. With novel schemes of experimental setup and data analysis \cite{Ross2011, Follett2016, Liu2019}, plasma parameters can be inferred with high accuracy, making Thomson scattering as key tool for quantitative study of high-energy-density physics. It becomes necessary that subtle effects must be included in the theory of Thomson scattering \cite{Zheng1997, Myatt1998, Zheng1999, Rozmus2000, Belyi2002, Tierney2003, Zheng2009, Palastro2010, Kozlowski2016, Rozmus2017, Belyi2018}. Many important physical processes have been experimentally investigated with this powerful 
technique \cite{Glenzer1996, Glenzer1999, Froula2007, Li2013, Rinderknecht2018, Henchen2018, Davies2019, Milder2020, Turnbull2020}. 

High-energy-density plasmas generated with high power laser pulses usually 
evolve rapidly with time because of great pressure gradient. 
In a modern experiment of optical Thomson scattering off laser-produced 
plasmas, scattered light waves are usually collected with an imaging 
system and are then relayed into a grating spectrometer coupled with a 
streak camera. With such kind of experimental setup, time-resolved 
Thomson scattering spectra are obtained, from which physical processes of 
interest are then inferred. However, to the best knowledge of the authors, 
the exact meaning of time-resolved spectrum obtained with above mentioned 
method is never clarified. Since Thomson scattering have been a very accuracy experiment tool for plasma physics, the meaning of the recorded signal should be carefully checked. In this article, we address this topic and discuss the relation between the recorded signal and the dynamic form factor.

\section{Output of a grating spectrometer}

\begin{figure}[ptb]
\centering{ \includegraphics[width=120mm]{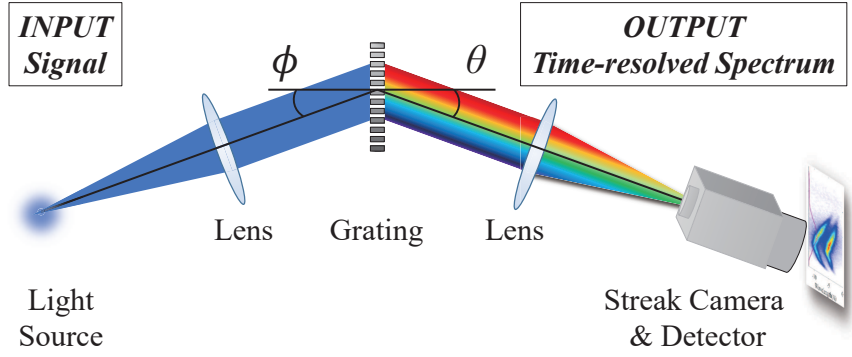} }
\caption{Schematic setup of time-resolved Thomson scattering measurement. }%
\label{setup}%
\end{figure}

We assume that a point light source is incident onto a grating after
collimated with an ideal lens. The diffracted light wave due to the grating is
then focused with another lens onto the slit of a streak camera. Shown in Fig.
\ref{setup} is the schematic setup, in which the incident angle is $\phi$ and
the diffraction angle is $\theta$. 

We describe the point source with a function $E(t)$. The output signal in the
spectral plane of the spectrometer is given by
\begin{equation}
I(t,\theta)=\left\vert \frac{1}{N}\sum_{k=-N/2}^{N/2-1}E(t-k\tau_{\theta
})\right\vert ^{2}, \label{output 1}%
\end{equation}
where $\tau_{\theta}$ is time delay between two diffracted light waves from
two adjacent groove lines of the grating,
\begin{equation}
\tau(\theta)=\frac{d}{c}(\sin\phi+\sin\theta). \label{time delay 1}%
\end{equation}
Here $d$ is the grating constant, $c$ is the light speed, and $N$ is the 
total number of the active groove lines of the grating. The output signal depends on the time $t$ as well as the diffraction angle $\theta$ which determines the frequency through the Bragg equation. Therefore, Eq. (\ref{output 1}) describes time-resolved spectrum.

With the introduction of the grating function $G$ defined as%
\begin{equation}
G(t)=\frac{1}{N}\sum_{n=-N/2}^{N/2-1}\delta(t-n\tau_{\theta}%
),\label{grating 1}%
\end{equation}
the output signal (\ref{output 1}) can be written as the square of the
convolution of the input signal and the grating function,%
\begin{equation}
I(t,\theta)=\left\vert \int_{-\infty}^{\infty}E(s)G(t-s)d\tau\right\vert
^{2}.\label{output 1A}%
\end{equation}
It is straightforward to show that the right hand side of Eq. (\ref{output 1A}%
) can be described with two Wigner functions \cite{Diels2006book},%
\begin{equation}
I(t,\theta)=\frac{1}{2\pi}\int_{-\infty}^{\infty}ds\int_{-\infty}^{\infty
}d\omega W_{G}(t-s,\omega)W_{E}(s,\omega).\label{output 1B}%
\end{equation}
Here the Wigner function for a function $F$ is defined as%
\begin{equation}
W_{F}(t,\omega)=\int_{-\infty}^{\infty}F^{\ast}(t-s/2)F(t+s/2)e^{-i\omega
s}ds.\label{Wigner 1}%
\end{equation}
As seen in Eq. (\ref{output 1B}), the time-resolved spectrum is the convolution of two Wigner functions: one is from the instrument of known, the other is from the signal of interest.  

Making the Fourier transformation of the Wigner functions with respect to the time variable,%
\begin{align*}
\hat{I}(\chi,\theta) &  =\int_{-\infty}^{\infty}I(t,\theta)e^{-i\chi t}dt,\\
\hat{W}_{F}(\chi,\omega) &  =\int_{-\infty}^{\infty}W_{F}(t,\omega)e^{-i\chi
t}dt,
\end{align*}
Eq. (\ref{output 1B}) can be simplified in the frequency domain,%
\begin{equation}
\hat{I}(\chi,\theta)=\int_{-\infty}^{\infty}\hat
{W}_{G}(\chi,\omega)\hat{W}_{E}(\chi,\omega)\frac{d\omega}{2\pi}.\label{output 1 in frequency}
\end{equation}

\section{Chirped pulse: an example}

%\subsection{Chirped pulse: analytical approach}

We are interested in the case that the frequency of the input signal is
dependent of time. As an example, we first consider the experiment that a
laser beam is scattered from a beam of uniformly accelerated electrons in
non-relativistic condition. We assume that the laser pulse is a gaussian and
that the velocity of the electron beam is given by
\begin{equation}
\mathbf{v}=\mathbf{a}t\text{ and }v\ll c. \label{electron velocity}%
\end{equation}
Here $c$ is the light speed in vacuum. In this case, the scattering process
can be approximated with the dipole radiation, and the signal can be described
with the following equation,
\begin{equation}
E(t)=E_{0}e^{-\alpha t^{2}/2+i\omega_{0}t+i\beta t^{2}/2},
\label{input signal 1}%
\end{equation}
where $\omega_{0}$ is the frequency of the laser probe, $\alpha^{-1/2}$ is its
pulse duration, and $\beta=\mathbf{k}\cdot\mathbf{a}$, and $\mathbf{k}$ is the
differential wave number of the scattering. It is easy to show that the Wigner
function of the input signal is then given by%
\begin{equation}
W_{E}(t,\omega)=2E_{0}^{2}\sqrt{\frac{\pi}{\alpha}}e^{-\alpha t^{2}%
-(\omega-\omega_{0}-\beta t)^{2}/\alpha}. \label{Wigner 2}%
\end{equation}
Its Fourier component is%
\begin{equation}
\hat{W}_{E}(\chi,\omega)=\frac{2\pi E_{0}^{2}}{\sqrt{\alpha^{2}+\beta^{2}}%
}e^{-\frac{\alpha\chi^{2}}{4(\alpha^{2}+\beta^{2})}-\frac{\alpha(\omega
-\omega_{0})^{2}}{\alpha^{2}+\beta^{2}}-i\frac{\beta(\omega-\omega_{0})\chi
}{\alpha^{2}+\beta^{2}}} \label{Wigner 2a}%
\end{equation}

The Fourier component of the Wigner function $\hat{W}_{G}(\chi,\omega)$ for
the grating can also be easily obtained,%
\begin{equation}
\hat{W}_{G}(\chi,\omega)=e^{i\chi\tau_{\theta}}\mathcal{G}(\omega
-\chi/2)\mathcal{G}(\omega+\chi/2), \label{Wigner grating 1}%
\end{equation}
where the function $\mathcal{G}(\omega)$ is defined as \cite{Wolf1999book}%
\begin{equation}
\mathcal{G}(\omega)=\frac{\sin(N\omega\tau_{\theta}/2)}{N\sin(\omega
\tau_{\theta}/2)}. \label{function G}%
\end{equation}
In the case of $N\gg1$, the function $\mathcal{G}(\omega)$ have a series of
sharp spikes locating at%
\begin{equation}
\omega\tau_{\theta}=2n\pi,\text{ where }n=0,\pm1,\pm2,\cdots.\label{Bragg equation}
\end{equation}
For the sake of analytical calculation, we make the approximation%
\begin{equation}
\mathcal{G}(\omega)\sim e^{-(\omega\tau_{\theta}-2\pi)/\sigma_N^{2}}, \text{ where }\sigma_N=0.7\pi/N.\label{approximation of function G}
\end{equation}
With this approximation, we have%
\begin{equation}
\hat{W}_{G}(\chi,\omega)=e^{i\chi\tau_{\theta}-\frac{(\omega\tau_{\theta}%
-2\pi)^{2}}{\sigma_N^{2}}-\frac{\chi^{2}\tau_{\theta}^{2}}{4\sigma_N^{2}}%
}.\label{approximation 1}%
\end{equation}

We introduce the diffraction angle shift $\Delta\theta$,
\[
\Delta\theta=\theta-\theta_{0}.
\]
where $\theta_0$ is defined as
\[
\frac{d}{\lambda_0} (\sin\phi+\sin\theta_0)=1, \text{ where }\lambda_0=\frac{2\pi c}{\omega_0}.
\]
After some basic calculations, we obtain the time-resolved spectrum of the
chirped pulse (\ref{input signal 1}) output from a grating spectrometer,
\begin{equation}
I(t,\Delta\theta)=\frac{(\sigma_N/\tau_{0})^{2}E_{0}^{2}}{\sqrt{\beta^{2}%
+(\alpha+(\sigma_N/\tau_{0})^{2})^{2}}}e^{-(a\omega_{0})^{2}\frac{\alpha
+(\sigma_N/\tau_{0})^{2}}{\beta^{2}+[a+(\sigma_N/\tau_{0})^{2}]^{2}}\left[
\Delta\theta+\frac{\beta}{a\omega_{0}}\frac{(\sigma_N/\tau_{0})^{2}}{\alpha
+(\sigma_N/\tau_{0})^{2}}t\right]  ^{2}-\frac{(\sigma_N/\tau_{0})^{2}\alpha t^{2}%
}{\alpha+(\sigma_N/\tau_{0})^{2}}},\label{output signal 1}%
\end{equation}
were $a$ and $\tau_0$ are defined as
\[
a=\frac{\cos\theta_0}{\sin\phi+\sin\theta_0},
\]
and 
\[
\tau_0=\tau_{\theta}(\theta_0).
\]
The result (\ref{output signal 1}) shows that at any given time $t$ the output signal has a maximum locating at the diffraction angle %
\[
\Delta\theta=-\frac{\beta}{a\omega_{0}}\frac{(\sigma_N/\tau_{0})^{2}}{\alpha
+(\sigma_N/\tau_{0})^{2}}t.
\]
The measured frequency shift $\Delta\omega=\omega-\omega_0$, which can be found through the Bragg equation (\ref{Bragg equation}), depends on the diffraction angle $\Delta\theta$ through the following equation,
\begin{equation}
\Delta \omega=-a\omega_{0}\Delta\theta.\label{measured frequency 1}
\end{equation}
The time-resolve spectrum of a chirped pulse output from a grating spectrometer is
\begin{equation}
I(t,\Delta\omega)=\frac{(\sigma_N/\tau_{0})^{2}E_{0}^{2}}{\sqrt{\beta^{2}%
+(\alpha+(\sigma_N/\tau_{0})^{2})^{2}}}e^{-\frac{\alpha+(\sigma_N/\tau_{0})^{2}%
}{\beta^{2}+[a+(\sigma_N/\tau_{0})^{2}]^{2}}\left[  \Delta\omega-\frac{(\sigma_N
/\tau_{0})^{2}}{\alpha+(\sigma_N/\tau_{0})^{2}}\beta t\right]  ^{2}%
-\frac{(\sigma_N/\tau_{0})^{2}\alpha t^{2}}{\alpha+(\sigma_N/\tau_{0})^{2}}%
}.\label{output signal 2}%
\end{equation}
As seen in Eq.~(\ref{output signal 2}), the measured frequency shift of the chirped pulse depends on time as
\[
\Delta\omega=\frac{(\sigma/\tau_{0})^{2}}{\alpha+(\sigma/\tau_{0})^{2}}\beta t.
\]
The changing rate of the measured frequency becomes smaller due to the dispersion of the grating. After integrating over the frequency $\Delta\omega$, we obtain the spectral-integrated signal,%
\[
I(t)\sim e^{-\frac{(\sigma_N/\tau_{0})^{2}\alpha t^{2}}{\alpha+(\sigma_N/\tau
_{0})^{2}}}.
\]
In comparison with the input signal, the output pulse becomes a little longer
due to the dispersion of the grating. In the case of
\[
\alpha\ll(\sigma_N/\tau_{0})^{2},
\]
the effects of the grating on the frequency and duration of the chirped pulse becomes negligible,
\[
I(t,\Delta\omega) = \frac{(\sigma_N/\tau_{0})^{2}E_{0}^{2}}{\sqrt{\beta^{2}%
+(\sigma_N/\tau_{0})^{4}}}e^{-\frac{(\sigma_N/\tau_{0})^{2}}{\beta^{2}%
+(\sigma_N/\tau_{0})^{4}}(\Delta\omega-\beta t)^{2}-\alpha t^{2}.}%
\]

\section{Time-resovled spectrum of Thomson scattering}

In an experiment of Thomson scattering off laser-produced plasmas, the
observed scattering spectra usually vary with time due to the temporal
evolutions of plasma parameters such as electron temperature and plasma 
flow velocity, etc. In the case that the incident wave is plane and 
monochromatic, the electric field of scattering waves from a plasma in 
non-relativistic case is given by \cite{Oberman1983}%
\begin{equation}
\mathbf{E}_{s}(\mathbf{R},t)=\frac{r_{e}\mathbf{n}\times(\mathbf{n}%
\times\mathbf{E}_{0})}{2R}\int_V d^{3}r\int d\tau\int\frac{d\omega_{s}}{2\pi
}\left[ e^{-i\omega_{s}(t-R/c)}  e^{i(\omega_{s}-\omega_{0})\tau-i(\mathbf{n}%
\omega_{s}/c-\mathbf{k}_{0})\cdot\mathbf{r}}+ \text{C.C.}\right]
\hat{n}_{e}(\mathbf{r},\tau).\label{scattering wave field 1}%
\end{equation}
Here $\mathbf{R}$ is the position of the objective lens of the scattering 
system, $\mathbf{n}=\mathbf{R}/R$ is the scattering direction, $r_{e}$ is 
the classical electron radius, $(\omega_{0},\mathbf{k}_{0})$ are the 
frequency and wave vector of the incident light wave, $\mathbf{E}_{0}$ is 
the amplitude of the incident wave, $\text{C.C.}$ denotes the complex 
conjugation 
term, and $\hat{n}_{e}(\mathbf{r},\tau)$ is the exact electron density
defined as
\begin{equation}
\hat{n}_{e}(\mathbf{r},\tau)=\sum_{j=1}^{N_{e}}\delta(\mathbf{r}%
-\mathbf{r}_{j}(\tau)).\label{fluctuating electron density 1}%
\end{equation}
The time-dependent part of the scattering wave field can be fully described
with the following function,%
\[
F(t)  =\int_V d^{3}r\int d\tau\int\frac{d\omega_{s}}{2\pi}\left(  e^{-i\omega_{s}%
t}e^{i\Delta\omega\tau-i\Delta\mathbf{k}\cdot\mathbf{r}}+ c.c. \right)  \hat{n}%
_{e}(\mathbf{r},\tau).
\]
where $\Delta\omega$ and $\Delta\mathbf{k}$ are given by
\begin{subequations}
\begin{align}
\Delta\omega & =\omega_{s}-\omega_{0},\label{differential frequency}\\
\Delta\mathbf{k}  & =\frac{\omega_{s}}{c}\mathbf{n}%
-\mathbf{k}_{0}. \label{differential wavevector}
\end{align}
\end{subequations}
The scattering waves are relayed into a grating spectrometer and then recorded
with a streak camera. The signal on the slit of the streak camera can be
written as%
\[
I(t,\theta)=\Delta\Omega\frac{cE_{0}^{2}}{16\pi}r_{e}^{2}\left[
1-(\mathbf{n}\cdot\mathbf{e}_{0})^{2}\right]  \int_{-\infty}^{\infty}%
ds\int_{-\infty}^{\infty}W_{G}(t-s,\omega)W_{F}(s,\omega)\frac{d\omega}{2\pi},
\]
where $\Delta\Omega$ is the solid angle of the collection system, and the Wigner
function $W_{F}(t,\omega)$ of $F(t)$ is now given by
\begin{align}
W_{F}(s,\omega) &  =\int dq\int_{V}d^{3}r\int d\tau\int\frac{d\omega_{s}}%
{2\pi}\int_{V}d^{3}r^{\prime}\int d\tau^{\prime}\int\frac{d\omega_{s}^{\prime
}}{2\pi}\left\langle \hat{n}_{e}(\mathbf{r},\tau)\hat{n}_{e}(\mathbf{r}%
^{\prime},\tau^{\prime})\right\rangle e^{-i\omega q}\nonumber\\
&  \times\left[  e^{-i\omega_{s}(s+q/2)}e^{i\Delta\omega\tau-i\Delta
\mathbf{k}\cdot\mathbf{r}} + \text{C.C.} \right]  \left[  e^{-i\omega_{s}^{\prime
}(s-q/2)}e^{i\Delta\omega^{\prime}\tau^{\prime}-i\Delta\mathbf{k}^{\prime
}\cdot\mathbf{r}^{\prime}}+\text{C.C.} \right]  .\label{time resolve 0}%
\end{align}
Here $\left\langle \hat{n}_{e}(\mathbf{r},\tau)\hat{n}_{e}(\mathbf{r}^{\prime
},\tau^{\prime})\right\rangle $ is the auto-correlation function of electron
density, and the notation $\left\langle \cdots\right\rangle $ means ensemble
average. 

The auto-correlation function $\left\langle \hat{n}_{e}(\mathbf{r},\tau
)\hat{n}_{e}(\mathbf{r}^{\prime},\tau^{\prime})\right\rangle $ in 
Eq.~(\ref{time resolve 0}) plays the central role in the theory of Thomson
scattering. For a stationary homogeneous plasma, the
auto-correlation function $\left\langle \hat{n}_{e}(\mathbf{r},\tau)\hat
{n}_{e}(\mathbf{r}^{\prime},\tau^{\prime})\right\rangle $ just depends on the
spatial and temporal dispalcements between the points $(\mathbf{r},\tau)$ and
$(\mathbf{r}^{\prime},\tau^{\prime})$. It is easy to show that the function
$W_{F}$ is independent of the time variable $s$, and is proportional to the dynamic form factor of the plasma, 
\begin{equation}
W_{F}(s,\omega)=4\pi n_{e}S(\Delta\mathbf{k},\Delta\omega).\label{stationary homogeneous 1}%
\end{equation}
Here $S(\Delta\mathbf{k},\Delta\omega)$ is the usual dynamic form factor of
the plasma, $n_{e}$ is the ensemble-averaged electron density. In this 
case, of course, the recorded scattering spectrum does not vary with time. 

When the plasma evolves slowly with space and time, the
auto-correlation function $\left\langle \hat{n}_{e}(\mathbf{r},\tau)\hat
{n}_{e}(\mathbf{r}^{\prime},\tau^{\prime})\right\rangle $ can be written as \cite{Belyi2018}%
\begin{equation}
\left\langle \hat{n}_{e}(\mathbf{r},\tau)\hat{n}_{e}(\mathbf{r}^{\prime}%
,\tau^{\prime})\right\rangle =\left\langle \hat{n}_{e}(\mathbf{r}^{\prime
}+\bm{\varrho},\tau^{\prime}+\eta)\hat{n}_{e}(\mathbf{r}^{\prime}%
,\tau^{\prime})\right\rangle .\label{auto-correlation 1}
\end{equation}
Now the auto-correlation function also depends on the space-time coordinates
$(\mathbf{r}^{\prime},\tau^{\prime})$ as well as the displacements
$(\bm{\varrho},\eta)$. We introduce a new function $Q(\mathbf{k}%
,\omega;\mathbf{r}^{\prime},\tau^{\prime})$ defined as
\begin{equation}
Q(\Delta\mathbf{k},\Delta\omega;\mathbf{r}^{\prime},\tau^{\prime})=\int%
_{V}d^{3}\varrho\int d\eta e^{i\Delta\omega\eta-i\Delta\mathbf{k}%
\cdot\bm{\varrho}}\left\langle \hat{n}_{e}(\mathbf{r}^{\prime
}+\bm{\varrho},\tau^{\prime}+\eta)\hat{n}_{e}(\mathbf{r}^{\prime}%
,\tau^{\prime})\right\rangle .\label{correlation function 1}%
\end{equation}
Generally, $Q(\Delta\mathbf{k},\Delta\omega;\mathbf{r}^{\prime},\tau^{\prime})$ is a complex function. Neglecting high frequency terms around $2\omega_{0}$, one can easily show that Eq.~(\ref{time resolve 0}) can be reduced into the following form,%
\[
W_{F}(s,\omega)=2\int\frac{d\omega_{s}}{2\pi}\int d^{3}\varrho^{\prime}\int
d\tau^{\prime}Q(\Delta\mathbf{k},\Delta\omega;\bm{\varrho}^{\prime},\tau^{\prime})e^{-i2(\omega_{s}-\omega
)s}e^{i2(\omega_{s}-\omega)\tau^{\prime}-i2[(\omega_{s}-\omega)/c]\mathbf{n}\cdot\bm{\varrho}^{\prime}}+ \text{C.C.},
\]
Introducing a new variable $\chi=2(\omega_{s}-\omega)$, we have
\begin{equation}
W_{F}(s,\omega)=\int\frac{d\chi}{2\pi}\int d^{3}\varrho^{\prime}\int
d\tau^{\prime}Q\left[  \Delta\mathbf{k}+(\chi/2c)\mathbf{n}%
,\Delta\omega +\frac{\chi}{2};\bm{\varrho}^{\prime},\tau^{\prime
}\right]  e^{-i\chi s}e^{i\chi\tau^{\prime}-i(\chi/c)\mathbf{n}\cdot
\bm{\varrho}^{\prime}}+ \text{C.C.}.\label{non-approximation}
\end{equation}
It is reasonable to assume that the function 
$Q(\Delta\mathbf{k},\Delta\omega;\bm{\varrho}^{\prime},\tau^{\prime})$ 
varies with $(\bm{\varrho}^{\prime},\tau^{\prime})$ in hydrodynamic 
scales. On the other hand, $\Delta\omega$ is the typical frequency of 
fluctuations measured with Thomson scattering. The following condition can 
be fulfilled in usual experiments,
\begin{equation}
\chi\ll\Delta\omega.\label{condition 1}%
\end{equation}
The typical frequency of hydrodynamic motions of the plasma is about $c_s/L_n$, where $c_s$ is the sound speed, and $L_n$ is the scale length of the plasma. Then we can make the estimation,
\begin{subequations}
	\label{condition 2}
\begin{align}
& (\chi/c)\mathbf{n}\cdot\bm{\varrho}^{\prime} \sim \frac{c_s}{c}, \label{condition 2A}\\
& \frac{\chi}{2c \Delta k} \sim 4\pi \sin(\theta_s) \frac{\lambda_0}{L_n} \frac{c_s}{c},\label{condition 2B}
\end{align}
\end{subequations}
where $\lambda_0$ is the wavelength of probe light, and $\theta_s$ is the scattering angle. In writing 
Eq.~(\ref{condition 2A}), we already assume that the size of the 
scattering volume is in the same order of the scale length, a condition usually satisfied in an experiment of laser-produced plasma.   
Since sound speed in a plasma is usually much slower than the light speed, we can neglect $\chi/c$ in Eq.~(\ref{non-approximation}). The function $W_{F}(s,\omega)$ can be
approximated as%
\begin{equation}
W_{F}(s,\omega)=2\int_V d^{3}r \operatorname{Re}Q\left(
\Delta\mathbf{k},\Delta\omega;\mathbf{r},s\right)  -\int_V
d^{3}r\frac{\partial^{2}}{\partial s\partial\Delta\omega
}\operatorname{Im}Q\left(  \Delta\mathbf{k},\Delta\omega;\mathbf{r},s\right)  .\label{time resolve 2}%
\end{equation}
If the function $Q(  \Delta\mathbf{k},\Delta\omega;\mathbf{r},s)  $
is real, Eq. (\ref{time resolve 2}) can be simplified,%
\begin{equation}
W_{F}(s,\omega)=2\int_{V}Q(  \Delta\mathbf{k},\Delta\omega;\mathbf{r}%
,s)  d^{3}r.\label{time resolve 3}%
\end{equation}
The recorded signal is given by%
\begin{equation}
I(t,\theta)=\Delta\Omega\frac{cE_{0}^{2}}{8\pi}r_{e}^{2}\left[  1-(\mathbf{n}%
\cdot\mathbf{e}_{0})^{2}\right]  \int_{-\infty}^{\infty}ds\int_{-\infty
}^{\infty}d\omega\int_{V}W_{G}(t-s,\omega)Q\left(  \Delta\mathbf{k}%
,\Delta\omega;\mathbf{r},s\right)  d^{3}r.\label{time resolve 4}%
\end{equation}

Equation (\ref{time resolve 4}) describes the time-resolved spectrum  
recorded with a streak camera, where the measured frequency is determined with Eq.~(\ref{measured frequency 1}). As indicated in the equation, the output signal \emph{is} the convolution of the grating and the spectral density of auto-correlation function of electron density. Guided with the result of Eq.~(\ref{stationary homogeneous 1}), we intuitively suggest that $Q\left(  \Delta\mathbf{k}%
,\Delta\omega;\mathbf{r},s\right)$ can be approximated as
\begin{equation}
Q(\Delta\mathbf{k},\Delta\omega;\mathbf{r},t) = n_e(\mathbf{r},t) S(\Delta\mathbf{k},\Delta\omega;\mathbf{r},t).\label{auto-correlation 2}
\end{equation}   
Here $S(\Delta\mathbf{k},\Delta\omega;\mathbf{r},t)$ is the dynamic form 
factor with inclusion of slow plasma evolution, which is recently carried out by V. V. Belyi \cite{Belyi2018}.

As a remark, it should be pointed out that the phase factor $(\chi/c)\mathbf{n}\cdot\bm{\varrho}^{\prime}$ in Eq.~(\ref{non-approximation}) may not always be negligible when a rapid process is studied. In a recent experiment performed by A. S. Davies \textit{et. al.}, picosecond thermodynamics in underdense plasmas was measured with Thomson scattering \cite{Davies2019}. In the experiment, $\chi\sim 10^{11}$ Hz, and the size of the scattering volume along the scattering direction is about $100$~$\mu$m. The phase factor $(\chi/c)\mathbf{n}\cdot\bm{\varrho}^{\prime}$ is about $0.1\pi$, a marginal value that can be neglected.    

\section{Summary}

In this article, we examine the meaning of the so-called time-resolved Thomson scattering spectrum that is usually encountered in experiments of laser-driven high-energy-density physics. When plasma evolves slowly, our result shows that the recorded signal is indeed the convolution of the response function of dispersion element and the spectral density of auto-correlation of electrons, i.e., Eqs. (\ref{time resolve 4}) and (\ref{auto-correlation 2}).

\begin{acknowledgments}
This work is supported by the National Key R \& D Projects (No.
2017YFA0403300), Science Challenge Project (No. TZ2016005), and the 
National Key Scientific Instrument Development Projects (No. ZDYZ2013-2).
\end{acknowledgments}

\bibliography{Thomson-References}

\end{document}